\input ./style/arxiv-general.cfg
\documentclass[MSNbibl,nameyear,dvips]{arxstspdf}
\makeatletter
   \@ifpackageloaded{graphicx}{}{\usepackage{graphicx}}
\makeatother
\usepackage{flushend}
\usepackage{stfloats}
\volume{30}
\issue{3}
\pubyear{2015}
\firstpage{423}
\lastpage{442}
\doi{10.1214/15-STS522}
\docsubty{FLA}


\begin{document}
\begin{frontmatter}

\title{A Conversation with Professor Tadeusz Cali\'{n}ski}
\runtitle{A Conversation with Professor Tadeusz Cali\'{n}ski}

\begin{aug}
\author[A]{\fnms{Anthony C.}~\snm{Atkinson}\corref{}\ead[label=e1]{a.c.atkinson@lse.ac.uk}}
\and
\author[B]{\fnms{Barbara}~\snm{Bogacka}\ead[label=e2]{b.bogacka@qmul.ac.uk}}
\runauthor{A.~C. Atkinson and B. Bogacka}

\affiliation{London School of Economics and Queen Mary, University of London}

\address[A]{Anthony C. Atkinson is Emeritus Professor, Department of
Statistics, London School of Economics, London WC2A 2AE, United Kingdom
\printead{e1}.}

\address[B]{Barbara Bogacka is Reader in Statistics, School of
Mathematical Sciences, Queen Mary University of London, London E1 4NS,
United Kingdom
\printead{e2}.}
\end{aug}

%
\begin{abstract}
Tadeusz Cali\'{n}ski was born in Pozna\'n, Poland in 1928. Despite the
absence of formal secondary eduction for Poles during the Second World
War, he entered the University of Pozna\'n in 1948, initially studying
agronomy and in later years mathematics. From 1953 to 1988 he taught
statistics, biometry and experimental design at the Agricultural
University of Pozna\'{n}. During this period he founded and developed
the Pozna\'n inter-university school of mathematical statistics and
biometry, which has become one of the most important schools of this
type in Poland and beyond. He has supervised 24 Ph.D. students, many of
whom are currently professors at a variety of universities. He is now
Professor Emeritus.

Among many awards, in 1995 Professor Cali\'{n}ski received the Order of
Polonia Restituta for his outstanding achievements in the fields of
Education and Science. In 2012 the Polish Statistical Society awarded
him The Jerzy Sp{\l}awa-Neyman Medal for his contribution to the
development of research in statistics in Poland. Professor Cali\'{n}ski
in addition has Doctoral Degrees \textit{honoris causa} from the
Agricultural University of Pozna\'{n} and the Warsaw University of Life
Sciences.

His research interests include mathematical statistics and biometry,
with applications to agriculture, natural sciences, biology and
genetics. He has published over 140 articles in scientific journals as
well as, with Sanpei Kageyama, two important books on the randomization
approach to the design and analysis of experiments.

He has been extremely active and successful in initiating and
contributing to fruitful international research cooperation between
Polish statisticians and biometricians and their colleagues in various
countries, particularly in the Netherlands, France, Italy, Great
Britain, Germany, Japan and Portugal. The conversations in addition
cover the history of biometry and experimental design in Poland and the
early influence of British statisticians.
\end{abstract}

%
\begin{keyword}
\kwd{Agricultural field trial}
\kwd{British Council}
\kwd{building a~department of statistics}
\kwd{communism}
\kwd{design of experiments}
\kwd{international collaboration}
\kwd{LINSTAT}
\kwd{Polish biometric school}
\kwd{Pozna\'{n}}
\kwd{randomization theory of experimental design}
\kwd{Rothamsted}
\kwd{Solidarno\'{s}\'{c}}
\end{keyword}
\end{frontmatter}

The conversation took place at the Pozna\'n University of Life Sciences
on 28.12.2012, 30.12.2013, 16.04.2014 and 15.07.2014.

%
\section{School and University Education}

\textbf{ACA:} You were born in 1928 so you should still have been
at school during World War II. However, under the German occupation,
education was in German and only at a very basic level, so you did not
go to school. Let us start in 1945 when you resumed your education in a
Polish school and started to make choices that would affect your
career.

\begin{figure}

\includegraphics{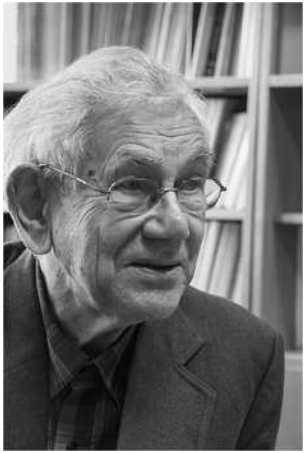}

  \caption{Tadeusz Cali\'nski, during the conversation (2014).}\label{fig1}
\end{figure}

\textbf{TC:} The fighting in Pozna\'{n} finished on 23 of February, 1945.
Immediately, some schools were trying to open to start the education of
Polish young people. On the first of March the Karol Marcinkowski
Gymnasium and Lyceum opened and I~passed the entry examination for the
first class of the gymnasium.

\textbf{BB:} But were there some underground schools in Pozna\'{n}
during the war?

\textbf{TC:} Not many. I~was lucky that my aunt was a teacher and after
work (in a medical equipment shop) I~went to her flat where she taught
me. Before the war I~had finished five classes out of six of primary
school. The sixth I~learned with my aunt and I~was admitted to the
first class of the gymnasium when the war finished. But because we were
so behind, we did two years in one, which was quite common in those
times. So in 1948 I~got the matura (secondary-school certificate).
Although I~liked mathematics, I~was persuaded by our Latin teacher to
major in humanities.

I~wanted to study forestry at the Faculty of Agronomy and Forestry of
Pozna\'{n} University, perhaps due to the influence of my uncle, a
forester. However, at that time, young people from the so-called
proletariat were accepted first, with extra points. Although I~passed
the entry examination, I~was not accepted because at that time my
father had a private business, so I~lost points. But I~was admitted in
October 1948 to study agronomy.

\textbf{BB:} May we ask what your parents were doing before the
war; what kind of profession?

\textbf{TC:} My father was a business representative for companies
producing cloth materials, and my mother was a qualified nurse. They
were not particularly interested in mathematics, but some relatives of
my maternal grandfather had mathematical talents. Also, one of his
cousins was an astronomer.

At that time the degree was in two parts: three and a half years for an
engineer and another year and a half for a master's degree. In February
1952 I~was awarded the Engineering Diploma. I~then chose to study for
the master's degree, under Professor Stefan Barbacki.

\begin{figure}

\includegraphics{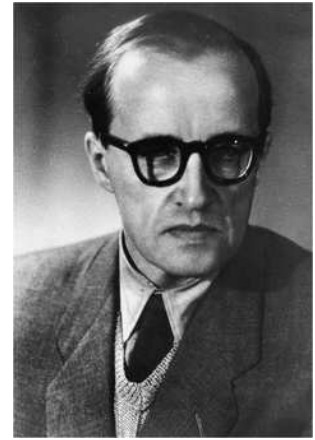}

\caption{Stefan Barbacki (1903--1979).}\label{fig2}
\end{figure}

I~received the master's degree in September 1953, although I~had been
working since February 1953. This happened because, in my final year,
Professor Barbacki gave an oral examination for his course on designing
experiments and their analysis. We were all together in one room. One
question was ``What is an interaction?'' There was a long silence and
then I~tried to explain how I~understood the term. Professor seemed to
be satisfied and after the examination he asked me ``Would you like to
work at my department?'' I~was very much astonished and do not remember
exactly how I~answered, but he offered me a job in his Department of
Genetics and Plant Breeding.

\section{Early Career and the Beginnings of Polish School of Biometry}

\textbf{TC:} The job was in the Institute of Husbandry, Fertilization and
Soil Science (Instytut Uprawy Nawo\.zenia i Gleboznawstwa, IUNG) at a
new experimental station in Babor\'{o}wko, not far from Pozna\'{n}. I~was employed by Professor Barbacki to help in conducting his field
experiments. This was my first professional experience.

\textbf{BB:} What kind of help? Did you have to go to the fields
themselves?

\textbf{TC:} Yes, I~worked there in the fields, although my first contact
with field experiments was during my master's degree at another
experimental research station in the Pozna\'n region and my master's
thesis was based on a field experiment. So, I~had already learned
something on how to conduct such experiments.

\textbf{BB:} In your master's degree, did you do some analysis of variance?

\textbf{TC:} Yes. Barbacki ran a lot of experiments and I~was involved in
some statistical analysis, working,
of course, not at a computer. I~had access to an
electro-mechanical calculator, rather like the one in the famous
photograph of Fisher. From 1951, the faculties of Agronomy and Forestry
were separate from the University and became an independent university
level Higher School of Agriculture (Wy\.zsza Szko{\l}a Rolnicza, WSR).
So, I~started my study at Pozna\'{n} University but finished it at WSR.

\textbf{BB:} In Poland, the need for proper statistical analyses of
agricultural trials was already advocated as early as 1906 by Edmund
Za{\l}\c{e}ski in his lecture on comparative experiments with different
varieties of sugar beet \textup{\citep{zaleski:1907}}. How did his work
influence the research of such people as Barbacki and Neyman?
Barbacki's book on the General Methodology of Field Experiments in
Outline was published in the same year \textup{[\citet{barbacki:1935}}, in
Polish\textup{]} as the famous book of Fisher on the Design of Experiments \textup{\citep{fisher:1935}}.

\begin{figure}

\includegraphics{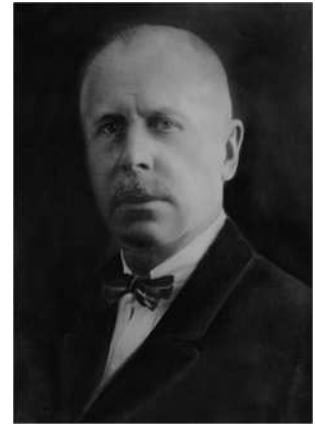}

\caption{Edmund Za{\l}\c{e}ski (1863--1932).}\label{fig3}
\end{figure}

\textbf{TC:} Za{\l}\c eski's lecture was published by the plant breeding
company of K. Buszczy\'nski and\break M. {\L}\c{a}\.zy\'nski, at which, from
1904, he was the head of research and plant breeding work. The paper is
regarded as the first systematic lecture on the methodology of
agricultural experiments with some application of probability theory.
It had three editions published in five languages. His ideas were
elaborated, from a mathematical point of view, by Jerzy Sp{\l
}awa-Neyman, whereas the practical applications were introduced by
several plant breeders, particularly Barbacki. In fact, in Poland,
earlier than in other European countries, new methods of improving the
precision of agricultural experiments were promoted.

\textbf{BB:} And how did the other schools, particularly of course
British, influence what was going on in Poland? For example, during
1935--1936 Barbacki had a stipend from the Rockefeller Foundation to visit
Fisher in London, which led to a trenchantly titled joint paper \textup{\citep
{AHG:AHG2138}} on the undesirability of systematic arrangements.

\textbf{TC:} Scientific relations between the British and Polish school
started in fact earlier. In September 1925 Neyman visited University
College, London; see \citet{reid:82}. A result of his time there was
the fundamental paper \citet{NP:28}.

\begin{figure}[b]

\includegraphics{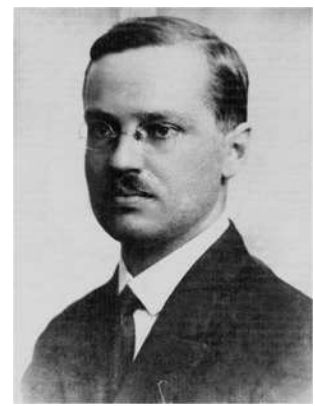}

\caption{Jerzy Sp{\l}awa Neyman (1894--1981).}\label{fig4}
\end{figure}

\textbf{BB:} After Barbacki's return from England he founded and
started to edit the Review of Agricultural Experimentation (Przegl\c{a}d
Do\'swiadczalnictwa Rolniczego), a monthly journal popularizing
statistical methods in agriculture. So, things seemed to be going well
in the 1930s, at least statistically.

\textbf{TC:} Yes, apart from Barbacki's book, Neyman, in particular,
published several papers on agricultural experimentation
(\citeauthor{Neyman:32}, \citeyear{Neyman:32,Neyman:34}).
However, the most important was
\citet{neyman1935} presented on March 28th, 1935, at a meeting of the Royal
Statistical Society.

{Two events at the end of the 1930s were important for Barbacki. In 1938
the procedure for his habilitation started at the Central College of
Agriculture (Szko{\l}a G{\l}\'owna Gospodarstwa Wiejskiego) in Warsaw.
In 1939 his next book, ``The Analysis of Variance in Problems of
Agricultural Experimentation,'' was published. Unfortunately, the whole
edition of the book was destroyed in September 1939 by the war events.}

\textbf{BB:} On September 1st, 1939, the Nazis invaded Poland. Soon
universities were closed and books burned. What is your understanding
of how it was in Pozna\'{n}?

\textbf{TC:} Pozna\'n was occupied by Germans from September 1939 to
February 1945. Pozna\'n University, as all other Polish universities,
was closed by the invaders almost immediately. Many professors and
other scientific employees were arrested and put in the Pozna\'n
extermination camp ``Konzentrationslager Fort VII Posen.'' Some others
were imprisoned or sent to various forced labor places in Germany. Most
of the remaining were resettled to the General Gouvernement (the Polish
territory occupied but not annexed to the State of Germany). Some of
those were then active in the underground University of the Western
Lands (Uniwersytet Ziem Zachodnich), started in 1940 in Warsaw.

Pozna\'n University reopened in early 1945, but with serious losses
among its teaching and research staff. The faculty of Agronomy and
Forestry itself lost 12 scientific workers, including 5 professors.

\textbf{BB:} Was Barbacki able to continue his scientific and
agricultural work?

\textbf{TC:} From 1925 to early 1945 Barbacki worked at the State Research
Institute of Rural Husbandry (Pa\'nstwowy Instytut Naukowy Gospodarstwa\break 
Wiejskiego) in Pu{\l}awy near Lublin. During the German occupation this
place was within the General Gouvernement. Fortunately, the institute
was not closed and continued its research work, of course under German
administration.

\textbf{BB:} In 2013 there was the 50th anniversary of the
Department of Mathematical and Statistical Methods at the Pozna\'n
University of Life Sciences. Barbacki was the founder of this
department, briefly run by Regina Elandt in the early sixties, but it
was you who made it grow and flourish. What were the beginnings of your
work there?

\begin{figure*}

\includegraphics{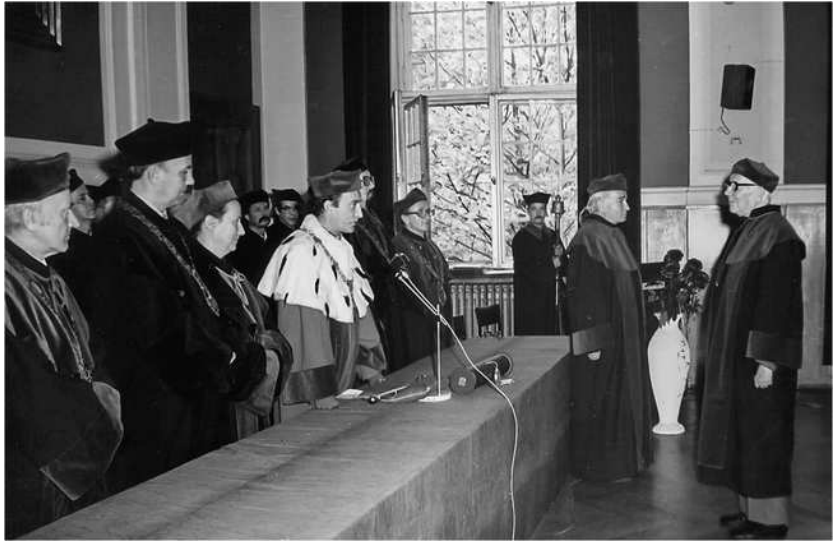}

\caption{W{\l}adys{\l}aw Orlicz (1903--1990) at his honorary degree
ceremony, Adam Mickiewicz University, Pozna\'{n}.}\label{fig5}
\end{figure*}

\textbf{TC:} When working with Barbacki from 1953 I~found that there is
something of mathematics in the statistical analysis of experimental
designs and data. Barbacki told me that if I~am so interested in
statistical methods, I~should learn more mathematics and undertake some
study of mathematics at the university. Fortunately for me, Professor
Orlicz, who was the Head of the Department of Mathematics at Pozna\'{n}
University, was a friend of Barbacki, and they both prepared a special
program for me.

\textbf{BB:} It means you had an individual program of study?

\textbf{TC:} Yes, I~had such a program for three years, but it was not for
a degree.
They decided which lectures I~should attend, which examinations I~should sit. This study was not full time, as I~was also working at
Barbacki's department where I~was associated with its section of
Experimental Design and Biometry.

\textbf{BB:} How was it to work with Barbacki?

\textbf{TC:} He was always very serious. He was the kind of person who
looked after the development of his people and he was very much
interested in what they were doing in science. He gave them appreciable
freedom in their research. He was not ``conducting'' them, rather he was
interested in what they were doing.

\textbf{BB:} This is probably the best way of supervising.

\textbf{TC:} Yes. He was also very much interested in people's wellbeing,
that they had enough money to support themselves. The salaries were low
and he tried to find extra jobs for his assistants; I~had an extra part
time position at the Department of Plant Genetics, Polish Academy of Sciences.

After World War I~there were two particularly important agricultural
research institutes in Poland. One was in Bydgoszcz, where Neyman
started his research work; the other was in Pu{\l}awy, with which
Barbacki was associated. It might be interesting to note that this
institute started in 1862 in the Russian part of the partition of Poland.

\textbf{ACA:} Was Za{\l}\c{e}ski involved in this institute?

\textbf{TC:} No. Before WWI Za{\l}\c{e}ski was a professor at the
Agricultural Academy in Dublany, one of the first Polish-language
schools of this kind, founded in 1858. It was near Lw\'ow (now Lviv in
Ukraine), at that time within the Polish area under the Austrian
Empire. Za{\l}\c{e}ski made enormous and innovatory methodological
achievements which he then compiled in his textbook \citet{zaleski:1927}.

\textbf{BB:} And he was the supervisor of Barbacki?

\textbf{TC:} Yes, after WWI Za{\l}\c{e}ski moved to Cracow and became a
professor at the Jagiellonian University, where Barbacki was a student
in the years 1921--1925. At the end of Barbacki's study he became Za{\l
}\c{e}ski's assistant. Then, after receiving his master's degree, he
obtained a job at the Pu{\l}awy Institute. In 1929 he received his Ph.D.
in agricultural sciences bestowed by the Jagiellonian University.

Because of the war his habilitation at the Agricultural University in
Warsaw was broken. After the war he was habilitated at Pozna\'{n}
University in 1945. There he organized the Department of Experimental
Design and Biometry---the first of this type in Poland. In 1951, when
the Faculties of Agronomy and Forestry had become the above-mentioned
WSR, he ran the Department of Genetics and Plant Breeding.

\section{Visits to the United Kingdom}

\textbf{ACA:} Could you tell us something about your visits to the
UK and their effect on your scientific development?

\textbf{TC:} My Ph.D. thesis in 1961 was ``On the Application of Analysis of
Variance to the Results of a Series of Varietal Experiments.'' My
supervisor, Barbacki, then decided that I~should spend some time
abroad. Actually, my first travel abroad was in 1960 for a mathematical
conference in Budapest. Regina Elandt (later Elandt-Johnson) had been
invited, but she was already traveling and suggested that I~take part
in that conference instead of her.

In 1964, due to Barbacki, I~was awarded a scholarship by the Polish
government to go abroad. I~became an honorary research assistant in the
Department of Statistics at University College, London, where Professor
Maurice S. Bartlett was the Head of the Department. I~think Norman
Johnson helped me in this, as I~knew him from his visits to Pozna\'{n}
to see Regina Elandt.

I~arrived to London in January 1964---almost in time to attend the
wedding of Regina with Norman.

\textbf{BB:} Why ``almost''?

\textbf{TC:} I~was to be there on the first of January, and their wedding
was on the 4th or 5th. But, as so often in Poland at that time, there
was a delay with my passport and I~missed the wedding.
I~arrived to London on the 10th.

Professor Bartlett was my supervisor there. I~got a shared room in the
department. I~could attend lectures and had time to do research. One of
the courses I~attended was given by David Cox. It was for the ``evening
students'' at Birkbeck College. The lectures were on the experimental
design, very important for me.

\begin{figure*}

\includegraphics{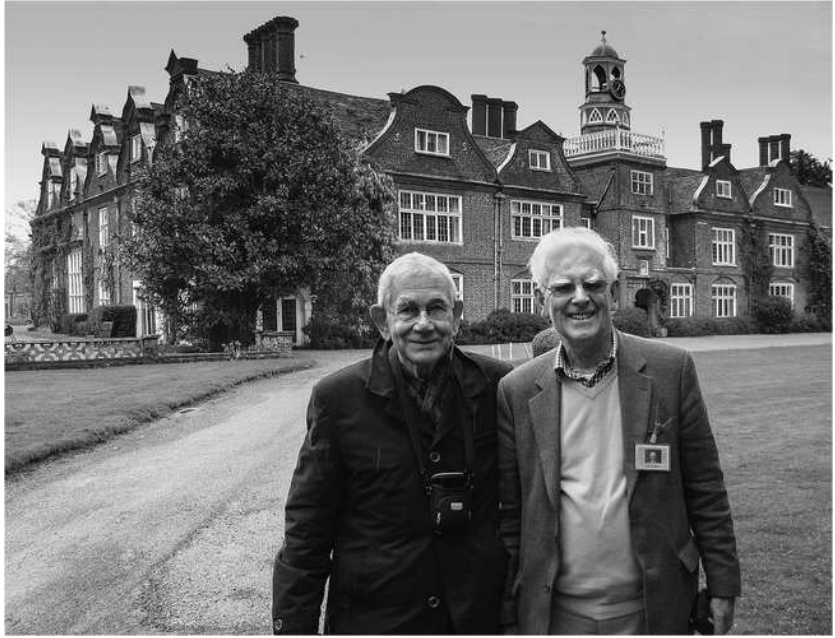}

\caption{T. Cali\'{n}ski and G. Ross in Rothamsted, 50 years after
Cali\'{n}ski's first visit there.}\label{fig6}
\end{figure*}

After some time Professor Bartlett told me ``as you are from an
agricultural university you should be interested in visiting the
Rothamsted Experimental Station.'' He suggested that I~write to Frank
Yates. So, I~wrote asking if it would be possible to spend some time in
Rothamsted. At that time, the English habit and the Post Office working
style were such that I~got the reply immediately, inviting me for an
interview. It was very interesting, but the interview was rather
difficult for me. It was difficult because Yates had a habit of keeping
his pipe in his mouth and it was hard for me to understand what he was
saying. Fortunately, Michael Healy was sitting next to him and he was
translating the Yates English to the BBC English.

The discussion was very interesting. He asked me what were my interests
in statistics. I~did not know his actual interests. Unfortunately, I~said I~was interested in mixed effects models. Then he asked, ``mixed
models, in which sense?''
I~made it even worse because I~replied ``in the Scheff\'{e} sense.'' He
became very unsatisfied because there was some conflict between him and
Scheff\'{e}, perhaps coming from an earlier meeting of the Royal
Statistical Society. Anyway, at the end of this interview he told me
``Yes, you can stay here as long as you wish, but remember, never
mention the name of Henry Scheff\'{e}.'' So, their conflict was quite
deep. I~started in April and stayed for half a year there.

I~had accommodation in the Manor House of Rothamsted. This was
excellent for me. The price was low, good food was served, very nice
place itself.

\textbf{BB:} Did you work there with some people?

\textbf{TC:} My supervisor was Desmond Patterson. I~was preparing my
habilitation ``On the Distributions of the $F$-type Statistics in the
Analysis of a Group of Experiments.''

\textbf{ACA:} Could you say something about this research? You were
using both the exact and approximate distributional results of
\textup{\citeauthor{box1954some1} (\citeyear{box1954some1,box1954some2})},
so it wasn't about design.

\textbf{TC:} Yes, the research concerned not designing but analyzing a
group of experiments (e.g., with crop varieties) conducted at different
places (environments). The main problem was the influence of the
heterogeneity of the error variance and of the interaction variance on
the distribution of the $F$-test statistics used in this type of
analysis. Some of Box's results were utilized in constructing
approximate tests. The discrepancies between the approximate and the
exact significance levels for the $F$-tests applied to experiments with
heterogeneous variances were investigated employing the Rothamsted
electronic computer.

\textbf{ACA:} And how did you spend your free time there?

\textbf{TC:} People at Rothamsted were very kind to foreign visitors; I~was
very often invited for private visits during weekends, sometimes by
Polish people who lived in Britain. But most frequently I~was invited
to their home by Rosemary and Gavin Ross who were then a young couple,
two years after their wedding. Gavin was in the Statistics section of
Rothamsted, with a particular interest in computing and simulation, and
he helped with my habilitation and with \citet{TC:66}.

There was also a Jewish lady, Miss Blanche Benzian, who often invited
me to visit her. What was really peculiar, that in London the only
people from the University staff who invited me to their homes were
Jewish. Although all members of the Department of Statistics were very
friendly, such private invitations I~was receiving exclusively from
English people of Jewish origin. At Rothamsted it was different, though
due to the hospitality of Blanche Benzian I~had the interesting
occasion to meet also some of her Jewish friends.

In September I~had some holiday and I~decided to go to Scotland. There
were two reasons that I~wanted to travel there. One was that I~wanted
to visit Professor David Finney. I~knew him well from the literature
and I~was very interested to meet him personally. As to the second
reason, there were some touring offers of British Council for foreign
students, and I~booked a trip along the Scottish Western Islands. After
the visit to Western Scotland I~went to Aberdeen and Professor Finney
presented me to his colleagues.

\textbf{ACA:} Did you attend any lectures or seminars there? I~remember, when he was at Edinburgh, Finney organized the seminars to be
at, I~think, five o'clock on Friday, to avoid his colleagues having
extended weekends.

You thank Finney at the end of \textup{\citet{TC:66}}. Did
you have any technical discussion with him about experimental design?
By the time you met he had already published his book on design \textup{\citep
{finney:55}} and was in Aberdeen to help establish a ``Rothamsted'' in Scotland.

\textbf{TC:} My discussions with David Finney, also with Desmond Patterson,
concerned mainly general problems that appear in conducting and
analyzing series of experiments, variety trials in particular.

\textbf{BB:} Walking in the mountains has always been your great
hobby. Did you have a chance to do some walking in the Scottish mountains?

\begin{figure*}

\includegraphics{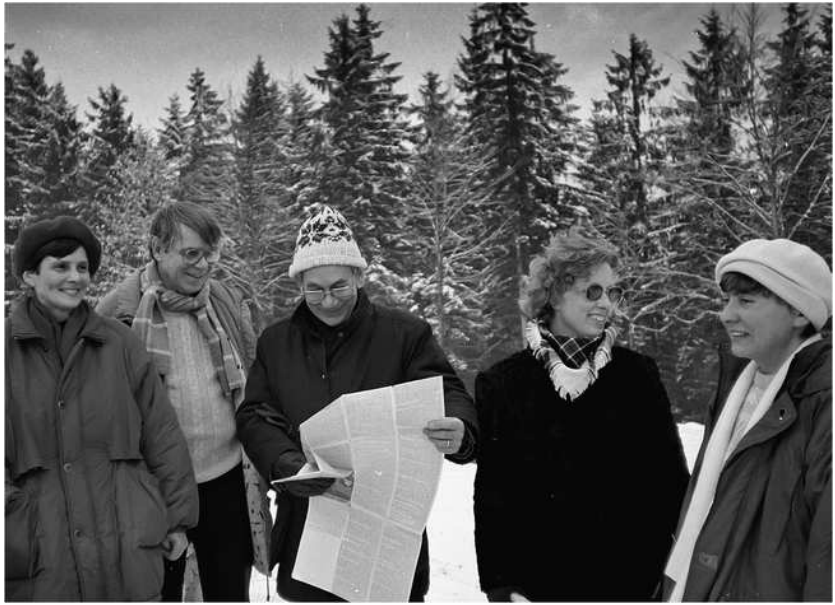}

\caption{During one of the Wis{\l}a conferences (1995).}\label{fig07}
\end{figure*}

\textbf{TC:} One of David Finney's colleagues was also Polish. He had a
camper van and we spent the weekend in the Cairngorm Mountains. We made
walks in the mountains, including mushroom picking, an activity not
very common among Scottish people. I~also visited the Highland Games at
Braemar. This was a great opportunity for me, as it was the first time
in my life that I~could see the Queen and her family.

\textbf{BB:} Let us come back to your visit to the UK. From
Scotland you came back to Rothamsted?

\textbf{TC:} Yes, then I~came back to Rothamsted and I~had another
opportunity of meeting people. It happened that William Cochran, the
author of a book on experimental design with Gertrude Cox \citep
{c+c:57}, was visiting Rothamsted at that time and also stayed in the
Manor House. One evening he told me about his experiences in Rothamsted
before the last war. He was working then with Frank Yates, with whom he
published several joint papers. Really, it was a great pleasure for me
to meet such an important person.

Of course, being in Britain I~had the opportunity to meet also some
other important persons. Particularly, I~would like to mention John A.
Nelder, whom I~had the chance to visit in the National Vegetable
Research Station at Wellesbourne, on a trip from Rothamsted.

So it was very interesting for me to be in Rothamsted. Then I~returned
for the last three months to London and continued my work there.

\textbf{ACA:} Did you talk to Bartlett? He was a very quiet man.

\textbf{TC:} Yes, but he was very friendly. Although he was extremely busy,
he was able to read my habilitation thesis and he gave me some
suggestions. I~remember when I~asked him to read it, he told me,
sitting at his desk, ``Do you see how many papers I~have to read? But
as you are leaving soon, I~will find one evening to read it.''

For the last month of my stay in London my wife visited me and we spent
December together. We had a week vacation organized by the British
Council and we took this opportunity to spend the week at the seashore,
at Broadstairs in Kent.

\textbf{ACA:} In December?

\textbf{TC:} Yes, again, there were many foreigners taking part in this
event since it was specially organized for foreign students. We also
had some friends in London, Polish people. So we spent Christmas Day
(December 25th, not 24th!) with them. Then we returned home, via Paris.
But I~decided that I~should go to Britain again, because I~liked the
country very much. Due to the help of Professor Finney, I~got a British
Council Scholarship for three months in 1969.

This time I~visited the University of Edinburgh (Finney's Department of
Statistics), University College Aberystwyth in Wales, and finally East
Malling Research Station in Kent. There I~started a very fruitful
collaboration with Clifford Pearce, leading to a paper in which we
introduced the notion of basic contrasts \citep{PEARCE01121974}.

\textbf{BB:} This work was a continuation of your previous results
\textup{\citep{calinski:71}}, which led to establishing a class of experimental
designs called ``C-designs.'' Were these the beginnings of your
interest in statistical properties of block designs?

\textbf{TC:} Yes, you could say so. Following these papers, several further
concepts and results concerning block designs were presented in
publications, for example, \citet{tc:93} and the joint papers on the
randomization model for experiments in block designs,
\citeauthor{cal+kag:96a} (\citeyear{cal+kag:96a,cal+kag:96b}).

\textbf{ACA:} Could you say something about the relationship of
this work to that of John Nelder on general balance?

\textbf{TC:} The joint work with Sanpei Kageyama followed some concepts of
Nelder, particularly those presented in \citet{nelder:54} and
\citeauthor{nelder1965analysis2}
(\citeyear{nelder1965analysis1,nelder1965analysis2}). Our first use of these
results was in \citet{cal+kag:91}.

During my second stay in Britain I~was very much interested in
attending the 37th Session of ISI in London in 1969, but this required
an extension of my visit; the British Council not only extended my
visit for two weeks, but paid for it as well. I~was able to attend (for
the first time) the ISI session. One interest in this meeting was that
I~wanted to see famous people whom I~knew from the literature,
particularly Anderson and Scheff\'{e}.

\textbf{ACA:} That was Ted Anderson?

\textbf{TC:} Yes, T. W. Anderson, the author of the book on multivariate
statistical analysis \citep{twa:84}. Some of the famous people I~had
already met, but it was impossible for me to meet those from the United
States. I~asked somebody how do I~recognize Henry Scheff\'{e}? I~was
particularly under the influence of his book on the analysis of
variance. Somebody told me ``if you see the most ugly looking
person, this will be Scheff\'{e}.'' This is so because he was a boxer or
wrestler and had a broken nose. So, finally, I~had the opportunity to
see Henry Scheff\'{e}.

\textbf{ACA:} Did he live up to the description?

\textbf{TC:} No, not so bad. During the meeting we also had some
opportunity to visit several places, such as certain London palaces. So
it was interesting scientifically but also very pleasant socially.


\section{Building the Department of Statistics in Pozna\'{n.}}

\textbf{BB:} Meeting and listening to the talks by all these famous
people, did they somehow influence your work?

\textbf{TC:} Since we wanted to learn multivariate analysis, when I~returned from Britain in 1965, we started a weekly seminar in Pozna\'
{n} based on the book by Anderson. Everybody had to present one of the
chapters, and in addition we had to solve and present the exercises.
The seminar was attended not only by people from our department, but
also Pozna\'{n} University (now called Adam Mickiewicz University---AMU) and from other institutes.

At the same time Witold Klonecki was giving the first course in Pozna\'
{n} on Mathematical Statistics at the Department of Mathematics, AMU.
He also ran a seminar for master's students. However, in 1966, Klonecki
was invited by Jerzy Neyman to go to Berkeley.
He was so fascinated by this that he stayed for two or three years. As
a result, the course in Mathematical Statistics ceased, at a time when
I~was very much interested to get new people to develop the department.

Although the standard of theoretical mathematics was very high at AMU,
Professor Orlicz wanted the Pozna\'n School of Mathematics to develop
in the direction of applications. In Klonecki's absence I~asked
Professor Orlicz to instruct one of his co-workers to continue the
seminar and the course. However,
Orlicz looked at me and said, ``You will do it!''
So I~started in September 1966. Orlicz told me, ``You will give the
course and prepare all the lectures.'' Also, he suggested that if there
were master's students interested in statistics I~should take care of
them, that is, I~should be supervising their master's theses.

\textbf{BB:} (reading the list) Altogether you supervised 53
master's degree students! I~see the last one is Pawe{\l} Krajewski in 1980.

\textbf{TC:} Some of these students got jobs at the AMU or the Technical
University in Pozna\'n, but many were employed in our department by
what is now called the Pozna\'n University of Life Sciences (PULS).

\textbf{BB:} Excellent! But why did the master's seminar stop in 1980?

\textbf{TC:}
It did not stop. Miros{\l}aw Krzy\'sko was habilitated in 1977 and
started a section of Probability and Mathematical Statistics within the
Department of Mathematics at the AMU where the seminar continued.

\textbf{BB:} Let us return to the earlier years of the seminar at
your department. Once you had your habilitation, you could formally
supervise a Ph.D.

\textbf{TC:} I~had got my habilitation in 1966. So, the seminar at our
department, from this point of view, was very fruitful. The year 1972
was particularly rich in Ph.D. theses. Altogether I~supervised 24 Ph.D. students.

\textbf{BB:} Five students working under your supervision defended
their theses in 1972. This is quite unusual, so many students finishing
in one year. How did you manage to achieve this?

\textbf{TC:} As I~said, the first Ph.D. theses were based on Anderson's book.
Such was that of Krzy\'{s}ko, who received his Ph.D. in 1971, as my first
Ph.D. student. The next few theses were also related to multivariate analysis.

\textbf{BB:} Did you continue the collaboration with Krzy\'{s}ko
and his new group?

\textbf{TC:} Certainly. We have been keeping contacts till now. In fact,
our last two joint papers both concerning canonical correlations are
\citet{calinski2005closed}, \citet{calinski2006comparison}.

\textbf{BB:} A different effect of your supervision came out from
the work of Micha{\l} Karo\'{n}ski. His seminar on Random Graphs is now
well established and brings collaborators from all over the world. Do
you know how his interest in this area started?

\begin{figure}

\includegraphics{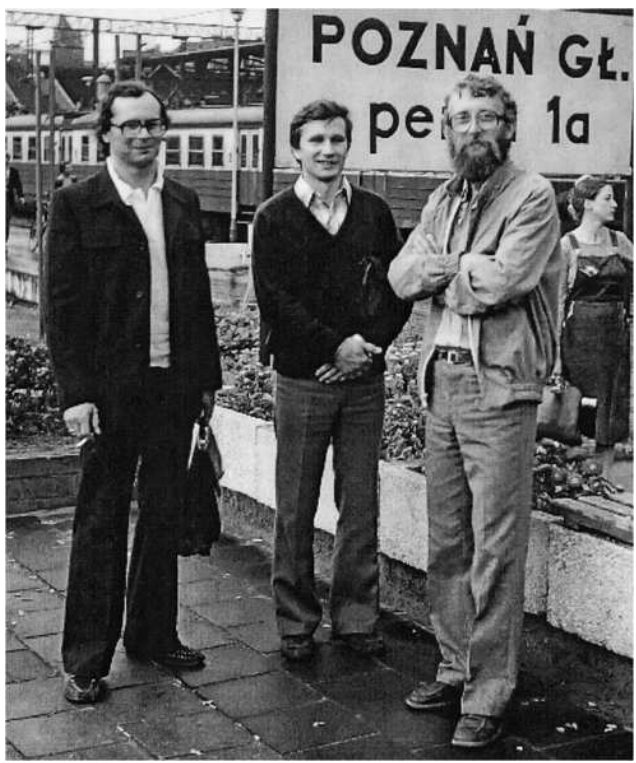}

\caption{J. Baksalary, P. Pordzik, R. Kala (1980, Photo by
G. H.~Styan).}\label{fig08}
\end{figure}

\textbf{TC:} He very early became interested in taxonomy. I~was his
supervisor for master's degree in 1968/1969. For this, he presented a
paper on Mahalanobis distances and their applications in taxonomy. His
Ph.D. thesis, presented in 1974, was on grouping multivariate populations
by using a simultaneous test procedure. This encouraged him to go
further, finally to random graphs.

\textbf{BB:} While Krzy\'sko and Karo\'{n}ski were actively
developing their groups, at your department the team of Baksalary and
Kala worked intensively on the theory of linear models, in particular,
aspects of linear algebra. Their joint work was published in such
journals as the Annals of Statistics, for example, \textup{\citet{BaksKal:81}}.
How did it happen that such a theoretical area was of so much interest
in your department?

\textbf{TC:} It started from a question that was raised at our seminar: how
to find bases of a matrix? I~suggested that perhaps somebody would
elaborate a useful method for this. One or two weeks later Baksalary,
and independently Kala, presented their solutions of this problem. So,
I~suggested that they sit down together and prepare a joint paper. In
fact, in the first ten years of their collaboration (1973--1982) they
published 41 joint papers in different journals, among them four in The
Annals of Statistics (the first in 1976 on Milliken's estimability
criterion). They both received their Ph.D.'s in 1975.

\textbf{BB:} Baksalary then moved to Zielona G\'{o}ra, where he
organized a very active group of mathematicians and statisticians.

\textbf{TC:}
Under Baksalary's rectorate (1990--1996) of the Tadeusz Kotarbi\'nski
Pedagogical University in Zielona G\'{o}ra, they started to organize
scientific meetings of mathematicians for the exchange of views on
various problems in linear algebra and mathematical statistics. It
seems that those meetings were very fruitful. They were called
``Konfrontacje Zielong\'orskie.''

\textbf{ACA:} What is the meaning of ``konfrontacje''? In English
it is rather aggressive.

\textbf{TC:} {No, here it is in the meaning of challenging ideas. It was
a yearly meeting, often in Zielona G\'ora. They invited people from
abroad, such as the Finn Simo Puntanen, to join the meetings. This was
an initiative of Baksalary. There is a special issue of \textit{Linear
Algebra and its Applications}, \textbf{410} (2005), devoted to Jerzy
Baksalary, in which a list of his co-authors is presented. There were
43 of them, including C. R. Rao.

\textbf{BB:} {Your work has spread to many places. One of them is
the Institute of Plant Genetics, Polish Academy of Sciences, where one
of your Ph.D. students, Zygmunt Kaczmarek, is working. Could you tell us
about the collaboration with this institute?}}

\textbf{TC:} My cooperation with this institute dates from 1956 and my
close collaboration with Zygmunt Kaczmarek from 1965. We have published
till now 51 joint papers (in the years 1968--2009). In 12 of these
publications, one of the co-authors is Pawe{\l} Krajewski, my master's
degree student and later Kaczmarek's Ph.D. student. He is now one of the
most active research workers in that institute. Among those 12 joint
publications, 6 concern our joint computer program SERGEN for analyzing
series of variety trials and plant genetic or breeding experiments.
This program has three editions and has been used by many plant
breeding and variety evaluation centers in Poland and also abroad, in
Great Britain in particular.

\textbf{ACA:} {When was the first Ph.D. thesis on experimental design?}

\textbf{TC:} Bronis{\l}aw Ceranka in 1972. This was the starting point of
his contribution to the theory and practice of experimental design.

\textbf{BB:} {Then, there was another student of yours, Wies{\l}aw
Pilarczyk, who worked for a long time at the Research Centre for
Cultivar Testing in S{\l}upia Wielka.}

\begin{figure*}

\includegraphics{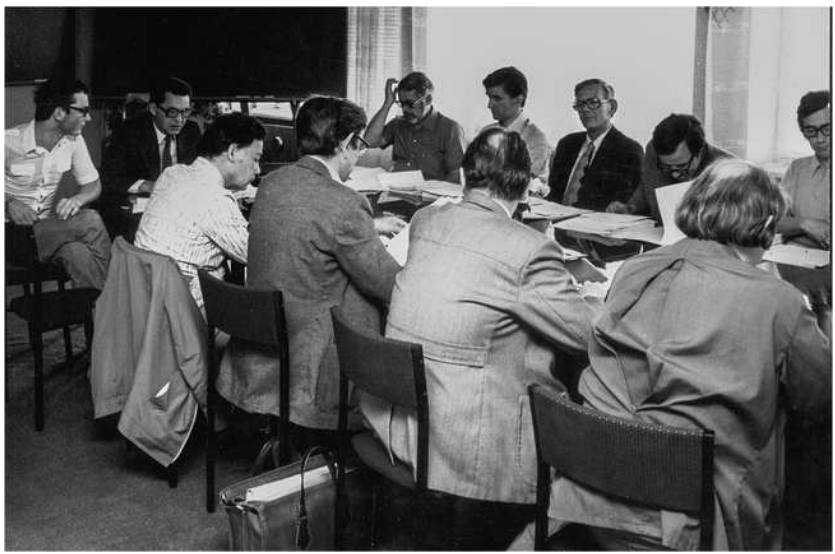}

\caption{First Working Seminar on Statistical Methods in Variety
Testing (1979).}\label{fig09}
\end{figure*}

\textbf{TC:} I~cooperated with this Research Centre (COBORU) from its
foundation in 1966. One of the results of this cooperation is the
series of Working Seminars on Statistical Methods in Variety Testing
organized jointly with our department. Each of these seminars was
attended by about twenty participants from several countries. Two very
important people attended the first seminar, in 1979: Professors Leo
Corsten and Desmond Patterson. Unfortunately, both of them died in 2013.

\textbf{BB:} {Your Ph.D. students have themselves supervised many Ph.D.s
over the years. These could be called your ``grandchildren'' and you
have done joint work with some of them.}

\textbf{TC:} From these 24 Ph.D.s we now have 12 with the title of professor.
Unfortunately, two of them have already died, Baksalary and Wagner. I~believe the exact number of grandchildren is currently 61. The most
important contributor to this is Micha{\l} Karo\'nski, who promoted 11
Ph.D. students, next are Miros{\l}aw Krzy\'sko with 9 promoted Ph.D.
students and Rados{\l}aw Kala with 8 promotions. Among them, there are
at least five foreigners: one each from Britain, Japan, Portugal and
two from Syria. There are also several great-grandchildren.

\textbf{BB:} {Over the years the Department of Mathematical and
Statistical Methods grew considerably and now it has over 30 academic
staff. But at the beginning it was rather small.}

\textbf{TC:} The department was founded in 1963 for Regina Elandt who was
so advanced in statistics [see \citet{elandt1964statystyka}] that
Barbacki thought she should start a new department of Mathematical
Statistics. However, she married Norman Johnson in January 1964 and
moved to Chapel Hill in July to join him.

When I~returned from Britain there were only two of us. But COBORU
started in 1966 and Eugeniusz Bilski left to become their director.
While collaboration with Bilski continued, I~was looking for new people
to join the department in Pozna\'n. Ceranka was the first such.

\textbf{BB:} {That was the time when you became the head of the department?}

\textbf{TC:} Actually, at the beginning there was no head. Although I~got
the habilitation in 1966, I~did not get the position immediately
afterwards. At that time, all positions at the university required
acceptance by a University unit of the communist party. Since I~was not
well seen by the local organization of the party, the first secretary
put the documents into a drawer of his desk. I~had to wait for two
years before I~got the position of Docent and then head of the
department. The party preferred to support their members. Fortunately,
there was no candidate from the party, so eventually I~became the head
of the department.

\textbf{ACA:}
{This was a kind of a pattern; some other people's careers were
delayed as well.}

\textbf{TC:}
Yes, for example, Kala had to wait a long time for the title of
Professor. His application papers, already supported by the Faculty,
likewise had to go to Warsaw for approval. Kala's habilitation was in
1981, but he was involved in Solidarno\'s\'c and so was punished by the
party. Kala only became a Professor in 1990. The eighties were the
worst years in our academic (and not only) life.

\textbf{BB:} {Even so, there was active research in the theory of
experiments, the theory of estimability in linear models, multivariate
analysis and matrix algebra and linear spaces. This is a wide spectrum
of interests for one department.}

\textbf{TC:} You have to remember that there was intensive cooperation with
people from other academic or research institutions, to some extent
under the influence of our joint weekly seminar.

\section{Statistical Education in Polish Universities}

\textbf{ACA:} {The recent history of Poland breaks very naturally
into decades. One of your activities (I~don't know how much time it
took) in the 1970s was in encouraging undergraduate education in
statistics. I~think this was initially a completely new idea in Poland?}

\textbf{TC:}
Again, this was due to Professor Orlicz who was at that time one of the
most important members of the Committee of Mathematical Sciences at the
Polish Academy of Sciences. One day he told me, ``I~have a duty for
you. I~will introduce you to our committee so that you become a member,
but you have to give a talk (expos\'{e}) explaining the importance of
mathematical statistics. We would perhaps then find some possibility of
supporting the development of the subject in Poland.'' I~demurred, but
he answered, ``You have to do it!''
Not accepting an ``invitation'' from Professor Orlicz was not an option.

\begin{figure}

\includegraphics{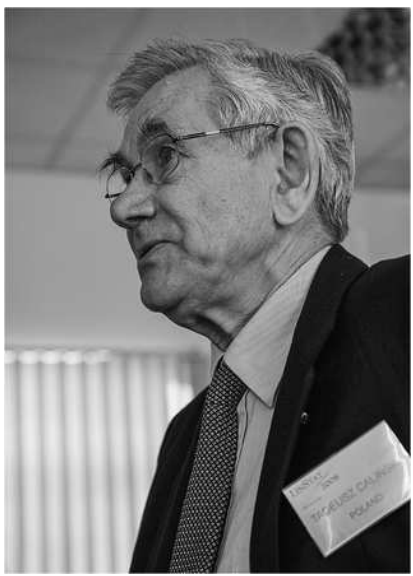}

\caption{Tadeusz Cali\'nski at LINSTAT 2008.}\label{fig10}
\end{figure}

So, I~joined this committee and I~gave an expos\'{e} on problems of
mathematical statistics, urging support for its development, which was
favorably received. A few months later, a commission was created for
development of mathematical statistics in Poland. I~became a
vice-chairman; J\'{o}zef {\L}ukaszewicz was the first chairman, as he
was from Wroc{\l}aw, where some applications of statistics were more
developed than in Pozna\'n. Witold Klonecki was also a member.

One of the ways of developing mathematical statistics in Poland was to
organize inter-university conferences. The first was in 1973 in Wis{\l
}a. These conferences were a very good platform for contacts with
people from various places in Poland and for inviting people from
abroad; one year it was a Polish national meeting and the other year it
was international. The \mbox{Commission} continues to work and the conferences
continue to take place.

\textbf{BB:} {Some of these international conferences were the
origin of the LINSTAT series. In fact, you organized the first one of
the LINSTAT series in Pozna\'n in 1984, at quite a difficult time in
Poland, briefly after Martial Law was withdrawn. I~remember that time
of political unrest and economical shortages, contrasted with the
high-spirited atmosphere at the conference and the lively discussions
in the conference room. LINSTAT became a regular conference, a European
meeting on mathematical statistics, largely devoted to applications of
linear algebra in statistics. The one in 2008 was to honor your 80th Birthday}

\textbf{TC:} This was really a great pleasure for me, particularly because
several of my old friends were able to participate in it; unfortunately
not all of them.

Another purpose of the Commission was to introduce lectures on
mathematical statistics at Polish universities, such as those Professor
Krzy\'sko was giving in Pozna\'n. They were also introduced in Warsaw,
Wroc{\l}aw and in many other universities at the departments of mathematics.

\textbf{BB:} {What was your role in the introduction of
statistics into the curriculum at the mathematics departments at Polish
universities, in Pozna\'n in particular?}

\textbf{TC:}
My role was to initiate the work, to organize the Commission's
activities, and to run meetings several times a year to discuss how to
support the idea of the development of mathematical statistics in
Poland. I~was chairman from 1975 to 1980, followed by Ryszard Zieli\'
nski (Warsaw), Miros{\l}aw Krzy\'sko (Pozna\'n) and Roman Zmy\'slony
(Zielona G\'ora). They and all other members contributed very much to
the activity of the Commission.

\textbf{BB:} {The document you presented in 1972 to the Committee of
Mathematical Sciences includes support for international collaboration,
help for libraries in the supply of international journals and books on
statistics and its applications and also assistance in obtaining funds
for research. Now, after more than 40 years of its work, how do you see
its achievements and its role for the future?}

\textbf{TC:} The Commission has fruitfully promoted statistical education
and research in Polish universities and research institutes. Now its
main activity is fostering cooperation between groups of mathematicians
interested in statistics within Poland and also with those from other
countries. The organization of national and international conferences
on mathematical statistics and its applications remains central. The
Pozna\'n group is quite active in this.

\section{International Collaboration}

\textbf{ACA:} {You not only followed Regina Elandt as Head of
Department, but you also followed her to North Carolina?}

\textbf{TC:} Yes, in 1977, from January to September, I~visited the
National Institute of the Environmental Health Sciences (NIEHS), its
Biometry Branch. I~had a very good time there, initially on my own,
latterly with my wife. For me being in the USA at that time in 1977 was
as to be on a different planet. The system was different, even to find,
for example, accommodation which, with the help of Norman Johnson, only
took a few hours. Also, the telephone was arranged in one day.
Everything was very easy, while in communist countries obstacles were
put in every step of our everyday life. We visited all the Eastern
States, from Florida to the Niagara Falls.

I~was accommodated at Chapel Hill, but the Institute was in the
so-called Research Triangle Park. The first institute, the Research
Triangle Institute, was founded in the sixties, the Statistics Research
Division being headed by Gertrude M. Cox. But when I~arrived there were
already more than 20 institutes or laboratories within that Park.

At the beginning I~had to travel to this place, about seven miles, but
there was no bus. So, a colleague was giving me a lift there every
morning. Before buying a car, I~attended a course for driving in
Durham. Fortunately the lady who was giving the course lived in Chapel
Hill, and she kindly drove me back home after the lessons. She was very
interested in the fact that I~was from a communist country. She asked
me, ``How is it possible that you can come to the United States from a
communist country?'' Well, I~said that this was possible. Then she
asked me, ``Did you have any problems with the FBI?'' I~said, ``Until
now, not.'' Then I~got an instruction from her: ``If you have any
problems with them, please let me know, because all of them have got
their driving licences at my course.''
But, actually, I~did not have any problems with that agency.\footnote{
For other comments from a visiting statistician on driving in North
Carolina see \citet{box:2013}, page~72.}

When my wife joined me, I~bought a small car, a~Volkswagen, which I~bought from a student for \$700. Then I~sold it, just before leaving,
for \$650. So, I~had a car for about half a year for 50 dollars.

We visited various places, mainly for scientific reasons. For example,
I~was invited by Professor William Mendenhall to the University of
Florida, in Gainesville, to give a seminar. I~was also attending the
Joint Statistical Meetings in Chicago, where I~gave a talk. Also, for
the first time in my life, I~saw Professor Kempthorne. I~found out that
he was not only a very good scientist, but also a very good dancer.
There was a conference dinner and after the dinner somebody said, ``Now
we have time for dancing.'' The first, who immediately asked a lady to
dance, was Oscar Kempthorne. They were dancing and we were all looking
at them, as very good dancers.

\textbf{BB:} {During your stay there did you undertake a special
research topic or collaborate with some people?}

\textbf{TC:}
In addition to Gainesville and Chicago, I~gave two seminars at the
University of Chapel Hill. I~also continued some written cooperation
with Clifford Pearce.

\textbf{ACA:}
{Did you have a delay in the mail? I~collaborated with Fedorov
in the Soviet Union and they put the letters in the bottom drawer for
three weeks.}

\textbf{TC:}
Yes, the contact from Poland was slow and in the US the letters to and
from Pearce came in a few days. I~was also preparing some work for
analyzing series of variety trials, in particular, in connection with
variety by environment interaction. Shortly after returning to Poland,
I~was invited to a conference in Italy and there I~presented the
results obtained in North Carolina.

\textbf{ACA:} {Before your visit to the United States, you were
already involved in collaborations with scientists in other countries.
How were these collaborations abroad funded? One of the features of
communist states was that they were always short of ``hard'' currency.}

\textbf{TC:}
There were various possibilities. In the early seventies some people
from our university were able to get scholarships from the Margrabina
Umiastowska Foundation. This was not controlled by the government but
by Polish people in exile, although somehow it was possible to use such
scholarships (perhaps some people from the Ministry were also
interested in traveling). Anyway, due to this foundation, I~received an
invitation (scholarship) in 1971 and spent one month in Italy.

This was a very busy time, visiting several people. The last was
Professor Giulio Alfredo Maccacaro, head of the Department of Biometry
and Medical Statistics at the University of Milan. In 1949--1950 he had
worked under Fisher at Cambridge. I~gave a talk on the methods of
multivariate analysis in biomedicine (later published as ``Metodi di
analisi multivariata in biomedicina''). Afterwards Professor Maccacaro
asked whether I~would like to come back to give a two week course. I~was very much astonished, but, of course, I~gave a positive answer. So,
my next stay in Italy was in Milan in May 1972.

\textbf{BB:} {I~remember, you were also visiting Italy in the
later years.}

\textbf{TC:}
Several times. After the course on Multivariate Analysis which I~gave
there, some of the geneticists became very interested. Mirella
Sari-Gorla was one of those who attended this course, also Alessandro
Camussi and Ercole Ottaviano. So, we started working together and there
have been several joint papers published as a result. Later Zygmunt
Kaczmarek joined us and also visited Milan several times and then also
Pawe{\l} Krajewski, both from the Institute of Plant Genetics, who
continued this collaboration. There is now a much larger group of
people, including scientists in the Netherlands, Ireland, Spain and the
United States, and many joint publications have appeared, for example,
the \textit{Science} paper \citet{Kaufmann02042010}.

\textbf{ACA:} {Mention of the work of Krajewski makes me realize
that we haven't so far at all mentioned your work on genetics.}

\textbf{TC:} As I~have said, at the beginning of my academic career I~worked at Barbacki's Department of Genetics and Plant Breeding. Also, I~cooperated with the Institute of Plant Genetics of the Polish Academy
of Sciences. My first participation in genetical research papers was in
\citeauthor{barbacki1978transgressions}
(\citeyear{barbacki1978transgressions,barbacki1978transgressions2}). Later,
some results of my cooperation with the Italian geneticists were
published in \citet{camussi1985genetic}, \citet{sari1997detection} and
\citet{calinski2000multivariate}.

My visits to Italy extended over several years. In 1983 I~was invited
by the Italian Region of the Biometric Society to give a course of
lectures in Gargnano, the northeast part of Italy, in a villa which
belonged to the University of Milan. They were organizing summer
courses there and I~gave lectures on Multiple Comparison Methods.

\begin{figure*}

\includegraphics{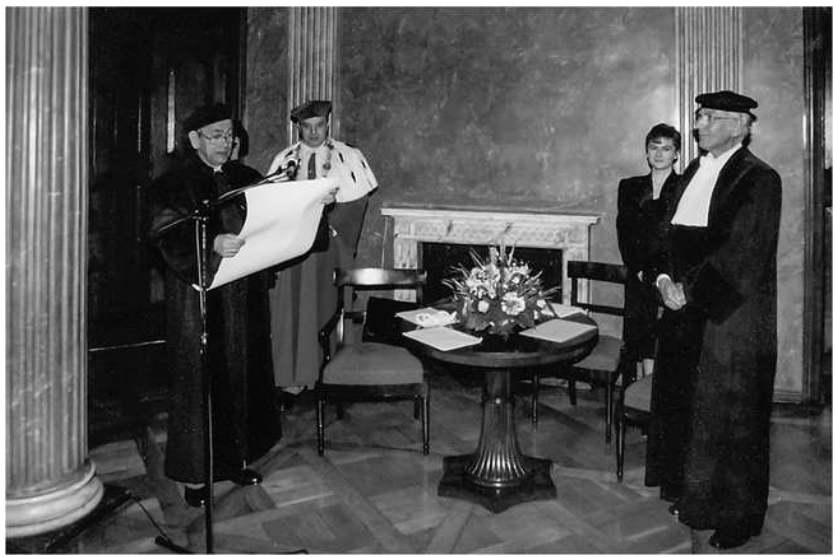}

\caption{Leo Corsten (1924--2013) during the Honorary Degree Ceremony,
1994.}\label{fig11}
\end{figure*}

Actually, I~attended these summer schools several times. Gargnano, on
the Lago di Garda, is historically very important because it was the
last headquarters of Mussolini. After the change of power in 1944,
Mussolini was arrested and held somewhere in the mountains, but the
Germans were able to free him. Most of Italy was already taken by the
allied forces, but the northern part was still under the German
occupation. After the war this house was assigned to the University of Milan.
Once, when I~was leaving, somebody asked me, ``Do you know whose room
you have been using?'' It was previously Mussolini's room!

The last visit to Italy was in 1997 for the Conference of the Italian
Region of the Biometric Society, where I~presented a paper based on
joint work with my French colleague Michel Lejeune \citep{TC+lej:98}. I~was also giving lectures in the south of Italy, at the University of Bari.

\textbf{BB:} {You also had a strong collaboration with people from
Wageningen in the Netherlands?}

\textbf{TC:} In 1970 the seventh International Biometrical Conference was
in Hanover. Clifford Pearce invited me to give a talk in his session on
experimental design. In that presentation I~made some references to
results of Leo Corsten. After my talk Leo approached me and said, ``I~am Leo Corsten to whom you were referring in your talk.'' So, that is
how it started. In 1975 I~visited Corsten's Department of Mathematics
at the Agricultural University of Wageningen, and in 1976 Leo Corsten
visited our Department in Pozna\'n. These visits led to an exchange
agreement between our universities. Rob Verdooren was the most frequent
visitor who also collaborated with COBORU. As a result of this
cooperation, I~have a joint paper with Leo Corsten published in
Biometrics \citep{calinski+cors:85} and Baksalary and Kala have a joint
paper with Corsten in Biometrika \citep{corsten1978reconciliation}. It
was a very fruitful cooperation and, of course, also it was a chance to
see a bit of the Netherlands.

The collaboration reached its climax in 1994 when Corsten got the
Honorary Doctorate from our University. It was an important event here.
In 1993 in Pozna\'n we had the second LINSTAT conference. In connection
with it a party was organized at which I~presented Corsten to our
Rector. He asked me later whether we should award Corsten an honorary
degree for the long-lasting cooperation.

\textbf{ACA:} {As well as visits abroad, international conferences
in Poland also provided a means of making new contacts.}

\begin{figure*}[b]

\includegraphics{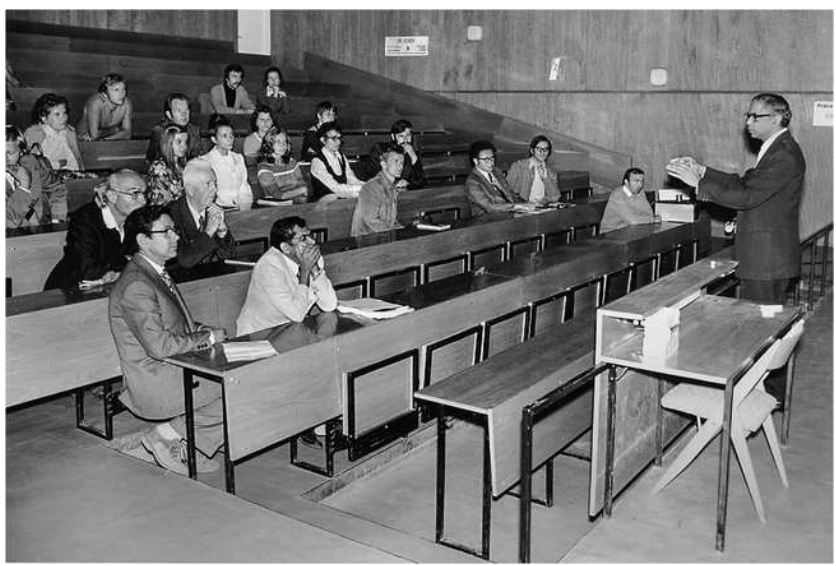}

\caption{C. R. Rao Speaking at the seminar in Pozna\'n, 1975.}\label{fig12}
\end{figure*}

\textbf{TC:} Certainly. The ISI Session in Warsaw in 1975 was an occasion
to invite several people to Pozna\'n. Among others, C. R. Rao was able
to visit us, also Parachuri R. Krishnaiah, Gavin Ross, Richard
Tomassone, Clifford Pearce, Dieter Rasch and Marie Jeanne Laurent-Duhamel.

As a result of the visit of R. Tomassone, head of the D\'{e}partement
de Biom\'etrie, Institut National de la Recherche Agronomique (INRA),
an agreement on cooperation with French biometricians was signed at the
ministerial level. I~visited INRA at Jouy-en-Josas in 1979, several
exchange visits followed and in 1982--1990 we jointly organized six
French--Polish Biometric Seminars, interchangeably in France or Poland.
Some joint research papers resulted from this cooperation.

\textbf{ACA:} {And how did you start your collaboration with
Kageyama, that led to the two books \textup{\citeauthor{cal+kag:2000}
(\citeyear{cal+kag:2000,cal+kag:2003})}?}

\textbf{TC:} I~first met Kageyama in 1983 when Klonecki and I~visited the
Indian Statistical Institute in connection with a conference in New
Delhi. My invitation to him to visit Pozna\'n initiated an exchange of
visits. Due to his support, I~received a research scholarship from the
Japan Society for the Promotion of Science so I~could visit him for
four weeks. It was a very well paid stipend and business class travel
expenses were also covered. However, with such support, I~had to work
very hard. During this one month, I~had to give seminars in seven
places at various institutes, including the Catholic Nanzan University
in Nagoya.

\textbf{ACA:} {Could they be the same talks?}

\textbf{TC:}
No, no. Different talks. Well, perhaps, some were a bit repeated.
Mainly, of course, the visit was connected to my joint work with Sanpey
Kageyama. So most of the time I~was staying at the Hiroshima
University. One of my lectures was for the Nagoya Chapter of the
Japan--Polish Society, where I~had to say something about Poland. So I~was very busy. Still we had time to work on the book.

\textbf{BB:}
{This is a book with a very broad coverage, so I~imagine it
required a lot of meetings and discussions. Do you still have contact
with him?}

\textbf{TC:}
Yes, in his last letter he said he was planning to retire completely in
2015. He has already retired from his university, but he is still
employed in a private university. In Japan, if you retire, you cannot
be employed again at the same university, but you can get a part-time
job at another one.

\textbf{BB:}
{The subject covered by the two books has also been actively
researched in the Department by several other people. The problem of
recovery of interblock information or of combining the estimators from
different block strata still needs more research, although several
papers were published on this issue. What is your view on the problem?}

\textbf{TC:} I~think the problem is basically solved, though some
extensions and practical applications would be welcome. Another paper
in this direction has been published recently \citep
{calinski2014combining}. Further results will probably be published by others.

\textbf{ACA:} {More recently, there has been an extensive
collaboration with Portugal.}

\textbf{TC:} In 2001 I~gave a two-week course for Ph.D. students on
experimental design in the Department of Mathematics at the New
University of Lisbon (Universidade Nova de Lisboa). I~was invited by
Professor Jo\~ao Tiago Mexia. In fact, I~was there five times, first in
2001 and the last in 2007, each time giving some courses.

In 2006 I~also took part in the 13th International Conference on the
Forum for Interdisciplinary Mathematics. This was in Tomar in Portugal,
connected with the occasion of an Honorary Doctorate for C. R. Rao.

\begin{figure}

\includegraphics{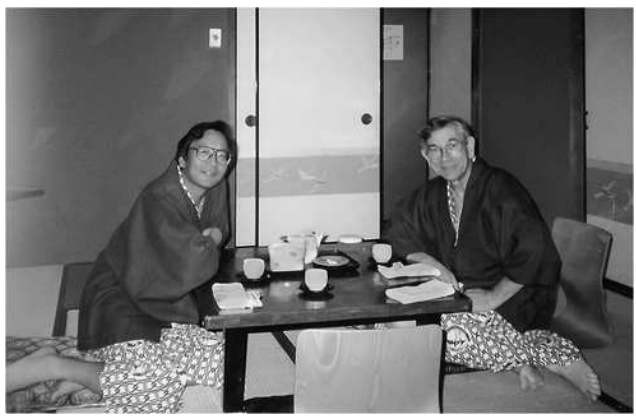}

\caption{Sanpey Kageyama and Tadeusz Cali\'nski.}\label{fig13}
\end{figure}

\section{``Solidarno\'s\'c'' and Martial Law}

\textbf{ACA:} {In 1980 there was the rise of Solidarno\'s\'c. Just
before martial law you became vice-rector for academic staff
development and international collaboration.}

\textbf{TC:} In 1981, for the first time, we had free election of the
authorities of the University. Previously the authorities had been
nominated by the Minister of Higher Education. In 1981 I~was elected as
a Vice-Rector for Staff Development and International Cooperation.
We started to work in September of that year. But as you know, in
December 1981 Martial Law was introduced in Poland. Then, at a meeting
of the Rector and Vice-Rectors, with the first secretary of the party
organization and a representative of the martial authorities,
a~military officer, we were told that according to their decision the
Rector had to resign. They thought that we would still work. Then I~said,
``We were elected as vice-rectors to this particular rector,
Professor Woj\-ciech Dzi\c ecio{\l}owski, and if he has to resign, we
also resign.'' So we all did. Thus, our work was very short, from
September until the beginning of January. We took part in the ceremony
of opening the new academic year in October, which was very impressive,
but our work soon stopped. Then, quickly, other people were appointed
to these positions.

However, I~was still a member of the University Senate and I~was still
attending its meetings. At one of these somebody from the Ministry of
Higher Education was present and he told me that they were dissatisfied
that I~resigned. This person was responsible for international contacts
and he secretly told me that he would help me to continue my
cooperation with France since our collaboration was part of an
agreement with France at the ministerial level. Therefore, despite
Martial Law, they allowed me to go to France in 1982, to take part in
the COMPSTAT Symposium in Toulouse, where I~gave an invited talk on
some problems in analyzing nonorthogonal designs. It was astonishing
that I~got a passport, as most people were not allowed to go abroad at
that time. But it was due to the support of the Ministry.

Later, we had the first French--Polish Biometric Seminar in Rennes,
which had already been accepted by the Ministry. So, I~and, this time,
my colleagues as well were allowed to go to France, despite the Martial
Law in Poland. I~traveled to the meeting in Toulouse by car together
with my wife, who also got a passport.
On the way we stopped at a car park somewhere in Germany, and somebody
noted that we were from Poland. He turned to others of his company and
said with a smile, ``Look, this is a brother of Wa{\l}\c{e}sa.''

\textbf{ACA:} {But, in general, travel abroad for extended periods
must have been very difficult in the 1980s.}

\textbf{TC:} Yes, it was quite difficult. Everything was under the control
of the first secretary. For example, in 1985 I~got an invitation to
Canada for half a year to give a course there. Although Martial Law was
over by then, I~did not get a passport. The Rector of our university
had a letter from the Ministry of Higher Education refusing me
permission to visit Canada due to ``negative environmental opinion.'' So
I~was not allowed to travel abroad for the whole year, even to
socialist countries.

However, as I~have said, we had a cooperation programme between the
Ministry of Agriculture of France and our Ministry. I~was responsible
for the section of Biometry. Some people in our Ministry were very much
interested to continue this cooperation. So, they supported me and at
the beginning of 1986 I~was able to go to France for half a year. But
when I~went to the passport office to get my passport somebody was
standing behind me and when my passport was put on the desk, that
person took it and asked me to go to the back of the office for an
``interview.'' He was from the secret police and said to me, ``We
finally decided that you can go.'' He also asked, ``How is it possible
that there are 32 people in your department and nobody belongs to the
party?'' I~said, ``I~am not a person who instructs my staff about
politics. You should ask this question to the first secretary of the
University's division of the party. Perhaps they do not make proper
propaganda.'' Then he told me that when I~am abroad I~should not contact
those Polish people who are against the Polish Democratic Republic and
so on. But this kind of instruction was quite common. The same
instructions which I~got before going to Britain.

\textbf{ACA:} {Did anybody ask you questions when you came back?}

\textbf{TC:} Usually, when we had instructions before leaving, they would
say, ``Remember when you are back home please contact us.'' But I~never did.

\section{Looking Back, Looking Forward}

\textbf{ACA:} {You retired in 1988. But since then you have
continued to be very much involved in statistical research and
collaboration. Looking at the citations of your publications, the paper
\textup{\citet{cal+hara:74}} is much more highly cited than any other of your
papers. What is it about, and how did it strike the jackpot?}

\textbf{TC:} Our paper was included in a comparison of 30 procedures for
determining the number of clusters in a data set. Our procedure
provided excellent results, taking the first place among all examined.
These results were published by \citet{millcluster:85}. Our procedure
was then incorporated into the SAS package.

\textbf{ACA:} {Thinking back over your scientific work, which
publications gave you most satisfaction?}

\textbf{TC:} First of all, the \textit{Biometrics} paper with discussion,
\citet{calinski:71}. It makes some suggestions for constructing useful
block designs that are simple in analysis and optimal in some sense for
practical application. These suggestions have been then advocated by
many authors, starting from \citet{saha1976calinski}. More recently, in
the book by \citet{raghavarao2005block}, Section~2.9 is devoted to
these patterns. Apart from this, the various papers published jointly
with my British, Dutch, French and Italian colleagues gave me much
satisfaction. Of course, also the joint works with Kageyama. I~have
also to mention that a subject of great interest, from the beginning of
my research work till now, has been the analysis of series of variety
trials. Most of my papers on this topic have been published jointly
with Stanis{\l}aw Czajka and Zygmunt Kaczmarek. The cooperation with
them, and later with some other colleagues, gave me great pleasure and
satisfaction. See, for example, our last paper, \citet{calinski2009mixed}.

\textbf{ACA:} {And which do you think have been unduly neglected?}

\textbf{TC:} I~have never thought about that.

\textbf{ACA:} {Of course, there is more to academic life than
publications. What activities have you found most rewarding?}

\textbf{TC:} Definitely the promotion of Ph.D. students and the scientific
support of young researchers. I~have been lucky to give lectures for
Ph.D. students and young researchers at various courses and conferences
organized in Poland and other countries. Even after my retirement in
1988, I~have had, eight times, the
pleasure to give such courses of lectures in several places abroad, in
Italy, Germany and Portugal. In general, international cooperation has
been giving me much satisfaction. I~am very glad that this type of
activity is continued by my younger colleagues in Pozna\'n,
particularly by Stanis{\l}aw Mejza, as president of the Polish
Biometric Society and organizer of the International Biometric
Colloquia in Poland; by Wies{\l}aw Pilarczyk, as a member of the
International Union for the Protection of New Varieties of Plants, and
coorganizer of the International Working Seminars on Statistical
Methods in Variety Testing; and by Augustyn Markiewicz, member of the
Pozna\'n Linear Algebra Group and coorganizer of the International
Workshops On Matrices and Statistics (IWMS), as well as of LINSTAT.
Also, some of my younger colleagues participate actively in various
international research projects. In particular, Pawe{\l} Krajewski is
very active in several international projects concerning plant
genetics, including the project Marie Curie ITN Epitraits, coordinated
by the University of Amsterdam. Results of these projects have been
published in leading journals. As far as our department is concerned, a
very active national and international cooperation has recently been
started by Idzi Siatkowski and his bioinformatics group. It includes
various research institutes in Poland, Switzerland, USA, Austria and
Germany. The first results of this cooperation were published in 2011,
for example, \citet{schmidt2011impact}.

\textbf{BB:} {All of this is largely due to your initiatives and
your actions.}

\textbf{TC:} Yes, but the initiatives were also supported and extended by
my co-workers, in the past particularly by Jerzy Baksalary. We were
very lucky here in Pozna\'n; despite the political situation, we were
able to establish very wide cooperation with other research institutes
in the world. I~am really very happy that some of my colleagues are now
very much involved in this.

\textbf{ACA:} {Looking to the future, do you have any hopes or
expectations for the future of our subject? Do you have any suggestions
for young people?}

\textbf{TC:} I~think that the future of mathematical statistics depends
very much on cooperation with other disciplines. So going back into
history, for example, to Neyman, he developed the theory of
mathematical statistics due to his strong contact with applied people.
His first papers were connected with agricultural experiments. He got
some ideas for estimation and for testing hypotheses by working on data
from agricultural experiments. You could say the same of Fisher and
many others, for example, Nelder.

So, if we really want to develop mathematical statistics, it is
important to encourage young people to be in touch with researchers
from other fields of science and industry and to be involved in the
analysis of data. This is well understood in our Department of
Mathematical and Statistical Methods, now headed by Anita Dobek (my Ph.D.
student of 1980).

\textbf{BB:} {Your wide international collaboration helped to
develop statistics in Poland. But you were also very active in
collaboration with other universities and institutes in Poland. These
activities have been recognized both by your honorary degree from the
Agricultural University of Warsaw and by many awards, the most recent
one being The Neyman Medal awarded by The Polish Statistical Society in
2012, the centenary year of the Society.}

\textbf{TC:} Yes, I~was quite involved in the organization of the Congress
of Polish Statistics at the occasion of 100th Anniversary of the Polish
Statistical Association. As to the Medal, it was awarded for those who
contributed to the development of statistical research in Poland.

\begin{figure}

\includegraphics{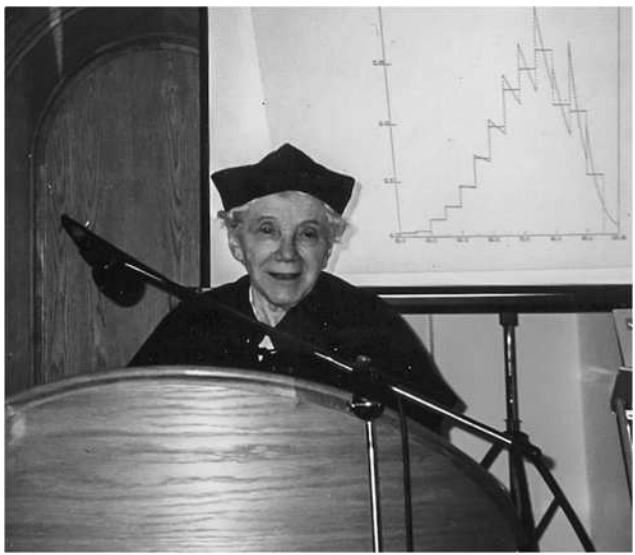}

\caption{R.~C. Elandt-Johnson at the Honorary Degree Award.}\label{fig14}
\end{figure}

\textbf{BB:} {Also, both Regina Elandt-Johnson and you were awarded
the Honorary Doctorate Degree from PULS, you in 1998 and she in 2001.
Were you her promoter for this award?}

\textbf{TC:} Yes, that is correct. It was a great honor for me.

\textbf{ACA:} {We have mentioned Neyman on several occasions. He, of
course, made much of his career outside Poland. Were there times when
you wished you had also moved abroad? Do you think you made the right
decision?}

\textbf{TC:} I~have never thought about leaving my country for another one.
I~liked to travel abroad, but I~did not have any temptation of staying
in a foreign country forever.

\textbf{ACA:} {And now to finish with a few questions about less
statistical matters, you said that there was not great interest in mathematics in
the earlier generations of your family. What about your children and
grandchildren? Any interest?}

\textbf{TC:} In pure mathematics not, in informatics yes. My son works for
IBM. So, some mathematical interest is somehow included. His son, Yann,
has an M.Sc. in Economics and, of course, he also attended lectures on
econometrics, which he liked very much.

\textbf{ACA:} {I~know you like traveling. You particularly travel
in East Poland and in Western Ukraine, is that right?}

\begin{figure}[b]

\includegraphics{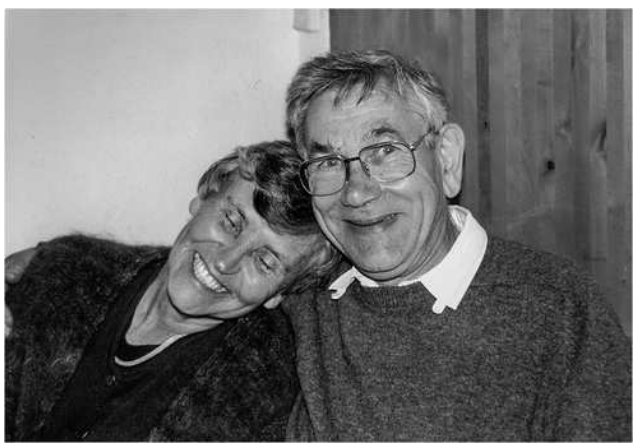}

\caption{Maria and Tadeusz Cali\'nski.}\label{fig15}
\end{figure}

\begin{figure}

\includegraphics{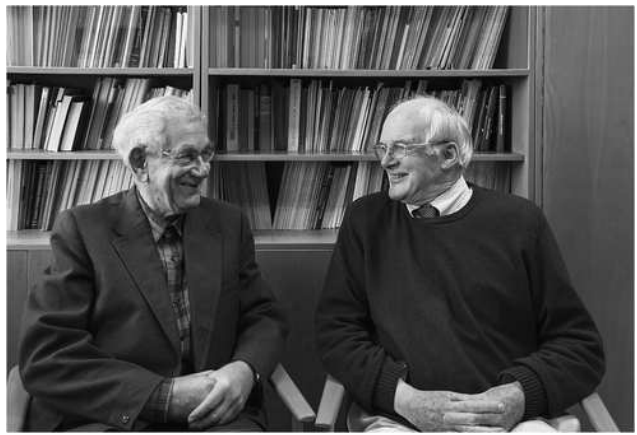}

\caption{Tadeusz Cali\'nski and Anthony Atkinson during the
conversation, 2014.}\label{fig16}
\end{figure}

\textbf{TC:} We have been traveling from the very beginning. But in the
nineties, when it was easier to travel wherever we wanted, some people
in our university started to organize excursions to those parts of the
previous Soviet Union which historically belonged to Poland, that is,
the eastern parts of that Poland which existed before the partitions at
the end of the eighteenth century. The first place we went to was Wilno
(Vilnius), as my wife was born there. We started in 1993 and visited
all our eastern neighbors: Lithuania, Latvia, Estonia, Belarus, Ukraine
and also Romania. The last was very interesting; we went southeast
through Ukraine to Romania and visited the northern part of the
country, Bucovina, where there are very unusual Christian orthodox
churches. It was near the town of Suczawa (Suceava), around which there
are several orthodox monasteries. They are very beautifully painted
both inside and outside. They originate from the 15th or 16th century
and are still in very good condition. Of course, this part did not
belong to Poland, except in the far past. These monasteries are
fortified. This part of Romania was always under some other country's
rule and it was not allowed to built fortresses around castles, but it
was allowed to do so around monasteries. One of the reasons for us to
go there was that there is still a Polish community living there, who
originally came to develop the salt mines.

\textbf{ACA:} {What about other interests of yours, for example, music?}

\textbf{TC:} I~very much like music, in particular, baroque music. In
Pozna\'n, every year we have a baroque festival. Last year it lasted
more than a week and there were about fourteen concerts. Of course, we
listen a lot to music on records. Every year in August there is also a
baroque festival called ``Music in Paradise'' organized in a monastery
Parady\.z, in Western Poland, which we very much like to attend.

\textbf{All:} {Thank you very much.}





\end{document}